# LUMINOSITY LIMITATIONS IN LINEAR COLLIDERS, BASED ON PLASMA ACCELERATION[*]

Valeri Lebedev[♦], Alexey Burov and Sergei Nagaitsev
*Fermi National Accelerator Laboratory, P.O. box 500, Batavia, IL60510*

*Abstract*
Particle acceleration in plasma creates a possibility of exceptionally high accelerating gradients and appears as a very attractive option for future linear electron-positron and/or γ–γ colliders. These high accelerating gradients were already demonstrated in a number of experiments. However, a linear collider requires exceptionally high beam brightness which still needs to be demonstrated. In this article we discuss major phenomena which limit the beam brightness of accelerated beam and, consequently, the collider luminosity.

## Introduction

In recent years two basic concepts for a linear electron-positron collider based on the acceleration in plasma were proposed and developed. They differ by the method of how the plasma-wave is excited. The first concept developed by the SLAC/UCLA group [1] is based on the plasma-wave excitation by a high-energy electron beam; and the second one, developed by LBNL [2, 3, 4] is based on the plasma excitation by a short pulse of laser radiation. In the case of a plasma excitation by an electron bunch, the length of a single accelerating section is much longer due to a much higher energy stored in the electron beam (compared to the energy sored in a laser pulse) and essentially negligible RF-dephasing with length. In the case of plasma excitation with a laser pulse, the RF dephasing is associated with the speed of a laser pulse being below the speed of light due to plasma permeability. Together dephasing and the energy stored in the laser pulse limit the length of acceleration to about 10 cm and the energy gain to about 10 GeV. Recently, a proposal to excite the plasma by a proton beam has also been actively discussed; and an experiment to validate such a concept is currently under construction at CERN [5]. The plasma excitation with a proton beam allows one to achieve the largest length of a single accelerating section which, potentially, could greatly simplify the accelerator. However, to our knowledge, nobody published a concept for a collider based on the proton-excited plasma acceleration. Below, we mostly concentrate on the acceleration of colliding beams. In this case, the method of plasma excitation does not play a significant role and with minor exceptions will not be discussed.

Plasma acceleration can be produced in two regimes. The first one is the quasi-linear regime, when the amplitude of the plasma electron density perturbation is smaller than the initial plasma electron density; and the second one, called the bubble regime, when the plasma electrons are completely blown away from the axis by the drive pulse. In this case, for a given plasma density the maximum accelerating gradient is achieved. As will be seen below, only the latter regime looks practical. In the case of electron bunch acceleration, this regime does not have any plasma electrons inside of the accelerated (witness) bunch and, consequently, does not suffer from repulsion of plasma electrons by the field of witness bunch, which degrades beam focusing in the bunch tail and results in emittance growth. As will be seen below, acceleration of positrons with brightness required for a collider is presently unfeasible for both regimes.

Both SLAC/UCLA and LBNL concepts suggest that they can achieve a collider operation with luminosity $10^{34}$ cm$^{-2}$s$^{-1}$ or above (see Table 1). However, a close examination of presented parameters shows that they are not completely consistent with limitations coming from fundamental principles of accelerator physics. To demonstrate it, the following limitations are considered below:

- Limitations on the beam emittance and momentum spread coming from the synchrotron radiation radiated in the final focus lenses/quads;
- Emittance growth due to multiple scattering in the plasma and/or residual gas;
- Pinching/expulsion of plasma electrons by electric field of positron/electron bunch;
- Pinching/expulsion of plasma ions (in the bubble regime) by electric field of electron/positron bunch;
- Bremsstrahlung on ions and impact ionization of

[*] Work supported by the US DOE under contract #DE-AC02-07CH11359
[♦] E-mail: val@fnal.gov

ions by the witness bunch electric field, if heavy ions are used to mitigate ion pinching by an electron bunch in the bubble regime;
- Transverse beam-breakup (BBU) instability due to the transverse wake-field excited in plasma.

Separately, each of above limitations can, in principle, be mitigated to an acceptable level. However they cannot be satisfied simultaneously within a set of ideas discussed up to present time. We knowingly do not consider technical limitations, which are very tight but, potentially, can be overcame with future developments in technology.

# 1. Final Focus and Beam Emittance Limitation

To be competitive with high energy proton colliders and, in particular, with the LHC, a lepton collider has to have the luminosity $\geq 10^{34}$ cm$^{-2}$s$^{-1}$. It is exceptionally challenging requirement, which demands very high beam brightness. As will be seen in the following sections, an increase of accelerated beam emittance would significantly simplify problems associated with acceleration of high brightness beam. However, a compensation of luminosity loss by tighter focusing in the interaction point (IP) is limited by many effects. In this section we consider the beam emittance and the momentum spread limitations caused by the final focus (FF) chromaticity and synchrotron radiation (SR) radiated in the FF lenses. Note that other effects, which are not considered inhere, may require even smaller emittances. However, the obtained values are already sufficiently small to switch on a number of other limitations discussed in the following sections. A more detailed discussion on final focusing limitations is presented in Ref. [6].

An increase of the beam emittance can be potentially compensated by an increase of focusing strength of the FF lenses resulting in a reduction of FF chromaticity and, consequently, a smaller beam size in the IP. Using plasma lenses instead of quadrupoles allows one to achieve focusing gradients orders of magnitude above what could be achieved using conventional technology. This could potentially greatly reduce effects of the FF chromaticity. However an increase of focusing gradient results in the SR from the FF lens. It limits the increase of the focusing strength.

To make a rough estimate, we assume that there is only one axially symmetric focusing lens[1] in the FF; the focusing length is equal to the lens length, $F=L_{lens}$; and the focusing chromaticity is suppressed by factor $\eta_F \gg 1$ with help of upstream optical elements. We also assume that in the case of flat beams, the vertical emittance and the IP beta-function are smaller than their values for the horizontal plane; and in the case of round beams, the horizontal and vertical emittances and the IP beta-functions are equal, respectively. Consequently, the vertical focusing chromaticity is larger than the horizontal one for flat beams, and the horizontal and vertical chromaticities are equal for round beams. Then, in the absence of chromaticity suppression, a particle with momentum deviation of $\Delta p/p$ is misfocused and its vertical offset in the IP is equal to:

$$\Delta y \approx F \sqrt{\frac{\varepsilon_y}{\beta_y^*}} \frac{\Delta p}{p} \ . \qquad (1)$$

---

[1] Axial symmetry implies that it is a plasma lens. Although with minor corrections this estimate is justified for quadrupole focusing.

Here $\varepsilon_y$ is the vertical single particle emittance (twice the particle action), and $\beta_y^*$ is the vertical beta-function in the IP. We also assume that the betatron phase is chosen to obtain the maximum particle displacement in the lens. Requiring this offset to be smaller than the maximum particle offset in the IP, $\sigma_y = \sqrt{\varepsilon_y \beta_y^*}$ (i.e. the offset of a particle with the same action but the betatron phase shifted by 90 deg.) one obtains limitation on the momentum deviation:

$$\frac{\Delta p}{p} < \eta_F \frac{\beta_y^*}{F} \ , \qquad (2)$$

where we also accounted for the suppression of focusing chromaticity of the final lens by a factor $\eta_F$. A finite value of $\eta_F$ is typically related to the higher order corrections and practical limitations of the correction scheme accuracy which we characterize by a single number. As one can see from Eq. (2) the limitation on the momentum deviation does not depend on $\varepsilon_y$.

The momentum change due to SR in the FF lens is[2]:

$$\left. \frac{\Delta p}{p} \right|_{SR} \approx \frac{2}{3} \frac{e^4 B^2 \gamma L_{lens}}{m_e^3 c^6} \approx \frac{2}{3} \frac{r_e \gamma^3}{F} \left( \frac{\varepsilon_x}{\beta_x^*} + \frac{\varepsilon_y}{\beta_y^*} \right) , \qquad (3)$$

where $\gamma$ is the particle Lorentz-factor, $e$ and $m_e$ are its charge and mass, $r_e$ is its classical radius, $B$ is the lens magnetic field at radius $r = \sqrt{x^2 + y^2}$ at particle location in the lens so that $B = Gr \approx GF\sqrt{\varepsilon_x/\beta_x^* + \varepsilon_y/\beta_y^*}$. In transition to the second half of the equation we also assumed that $1/F \approx eGL_{lens}/(mc^2\gamma)$ and $F=L_{lens}$. Equaling $\Delta p/p$ of Eqs. (2) and (3) for particles with $2\sigma$ amplitudes one obtains limitations on the normalized rms beam emittance,

$$\left. \varepsilon_{ny} \right|_{max} \equiv \varepsilon_y \gamma \leq \frac{1}{4} \sqrt[3]{\frac{3S^{*2}\eta_F}{4\pi^2 r_e R^2 R_\varepsilon}} \ , \qquad (4)$$

---

[2] Gaussian units are used through this article.

and the relative rms momentum spread

$$\left.\frac{\Delta p}{p}\right|_{\max} \leq \frac{\gamma}{F}\sqrt[3]{\frac{4r_e S^* \eta_F^2 R_\varepsilon}{3\pi R}} \quad . \tag{5}$$

Here $S^* = \pi\sigma_x^* \sigma_y^*$ is the effective beam cross-section at the IP, $R = \sigma_x^*/\sigma_y^* \geq 1$ is the ratio of beam sizes in the IP, and $R_\varepsilon = (1+\varepsilon_x \beta_y^*/\varepsilon_y \beta_x^*)/2$ is the parameter characterizing the ratio of beam sizes in the lens. Taking into account that $S^*$ is determined by the luminosity one can see from Eq. (4) that the limitation on the beam emittance does not depend on the lens focusing strength and has weak dependence on $\eta_F$, i.e. stronger focusing does not help to increase the beam emittance. However as one can see from Eq. (5) it allows one to accept a larger momentum spread.

For the luminosity of $2 \cdot 10^{34}$ cm$^{-2}$s$^{-1}$, bunch frequency of 15 kHz and the beam energy of 0.5 TeV one obtains $S^*$=314 nm$^2$. Assuming round beam ($R$=1, $R_\varepsilon$=1), a factor of 100 times suppression of the FF chromaticity ($\eta_F$ = 100) and the focal distance of $F$ = 1 m, one obtains from Eqs. (4) and (5) that $\varepsilon_{ny} \leq$ 1.6 µm and $\Delta p/p \leq 1.5 \cdot 10^{-2}$. Note that this estimate is still quite optimistic, and it will be significantly smaller, if quadrupoles are used instead of an axial symmetric lens.

Table 1: RMS Emittances and Momentum Spreads for Various Linear Collider Proposals

|  | ILC | CLIC | LPA | PWFA |
|---|---|---|---|---|
| Beam energy, TeV | 0.25 | 1.5 | 0.5 | 1.5 |
| Luminosity, $10^{34}$cm$^{-2}$s$^{-1}$ | 1.8 | 6 | 2 | 6.3 |
| Particle per bunch, $10^{10}$ | 2 | 0.37 | 0.4 | 1 |
| Bunch rep. rate, kHz | 6.5 | 15.6 | 15 | 10 |
| $\sigma_x$, nm | 474 | 40 | 10 | 194 |
| $\sigma_y$, nm | 6 | 1 | 10 | 1.1 |
| rms norm.h.emit., $\varepsilon_x$, µm | 10 | 0.66 | 0.1 | 10 |
| rms norm.v.emit., $\varepsilon_y$, µm | 0.035 | 0.020 | 0.1 | 0.035 |
| Rms mom. Spread, % | 0.1* | 0.35 | N/C | N/C |
| Rms bunch length, µm | 300 | 44 | 1 | 20 |
| IP size ratio, $R=\sigma_x/\sigma_y$ | 79 | 40 | 1 | 176 |
| $R_\varepsilon = (1+\varepsilon_x \beta_y^*/\varepsilon_y \beta_x^*)/2$ | 7 | 0.84 | 1 | 1.8 |
| Emit. margin $\varepsilon_{ny}\|_{\max}/\varepsilon_y$ | 12 | 4 | 16 | 2 |

* - Taken from Ref. [8]
N/C – not cited

Table 1 presents parameters for plasma based colliders [7] suggested by SLAC/UCLA (PWFA - plasma wake field accelerator) [1] and LBNL (LPA – laser plasma accelerator) [2,3,4]. Parameters for much more mature ILC and CLIC projects are also presented [7,8]. One can see that the beam emittance for an LPA is approximately an order of magnitude smaller than in the above estimate. At the first glance it looks as a significant margin. However pinching of plasma electrons by bunch electric field, discussed in the following sections, excludes possible use of a plasma lens at the required beam brightness. The use of quadrupoles leaves no margin for possible emittance increase for the LPA. The last line in Table 1 presents the ratio of emittance limitation of Eq. (4) to the rms emittance presented in the same table. As one can see the margins for different machines are not significantly different.

## 2. Particle Focusing and Acceleration in a Plasma

In plasma based accelerators, the thermal velocities of plasma electrons are much smaller than the velocities excited by plasma wave and the motion of ions can be neglected. In this case a plasma wave can be described by a hydro-dynamical approximation, where only electron motion is accounted:

$$\begin{cases} \dfrac{\partial n_e}{\partial t} + \text{div}(n_e \mathbf{v}_e) = 0, \\ \text{div}(\mathbf{E}) = -4\pi e(n_e - n_0), \\ \dfrac{\partial \mathbf{v}_e}{\partial t} + (\mathbf{v}_e \cdot \nabla)\mathbf{v}_e = -\dfrac{e}{m_e}\left(\mathbf{E} + \dfrac{1}{c}[\mathbf{v}_e \mathbf{B}]\right), \end{cases} \tag{6}$$

where $n_0$ is the unperturbed electron density, and the sign "minus" takes into account the negative charge of plasma electrons. Linearizing these equations, one obtains:

$$\begin{cases} \dfrac{\partial \tilde{n}}{\partial t} + n_0 \text{div}(\mathbf{v}_e) = 0, \\ \text{div}(\mathbf{E}) = -4\pi e \tilde{n}, \\ \dfrac{\partial \mathbf{v}_e}{\partial t} = -\dfrac{e}{m_e}\mathbf{E}. \end{cases} \tag{7}$$

where $\tilde{n} = n_e - n_0$ is the plasma density perturbation. Excluding $\mathbf{v}$ and $\mathbf{E}$ from the above equations one obtains:

$$\frac{\partial^2 \tilde{n}}{\partial t^2} + \omega_p^2 \tilde{n} = 0 \quad , \tag{8}$$

where $\omega_p = \sqrt{4\pi n_0 e^2/m}$ is the plasma frequency. Eq. (8) shows that in the linear regime the plasma oscillates at the plasma frequency, independently of the initial density perturbation.

To find conditions of applicability of the linearized Eqs. (7) we consider a one-dimensional solution with all values changing as $e^{i(\omega t - kz)}$. Then, one obtains from Eqs. (7) the following relationships:

$$\begin{cases} v_{e_z} = c\tilde{n}/n_0, \\ E = -E_0 \tilde{n}/n_0. \end{cases} \tag{9}$$

Here, we took into account that the plasma wave is propagating with the speed of light, $\omega = kc$, $\omega = \omega_p$, and

$$E_0 = \frac{4\pi e n_0}{k} \qquad (10)$$

is the electric field amplitude for a harmonic perturbation of plasma electrons density with 100% amplitude ($\tilde{n}/n_0 = 1$). As one can see, a density perturbation comparable to the initial plasma density results in velocities of plasma electrons comparable to the speed of light. Consequently, if the density perturbation is comparable to the initial density, the relativistic equations for plasma electron motion have to be used in Eqs. (6).

Performing a Fourier transform in time for the following Maxwell equation:

$$\text{rot }\mathbf{B} = \frac{4\pi \mathbf{j}}{c} + \frac{1}{c}\frac{\partial \mathbf{E}}{\partial t} \qquad (11)$$

and obtaining from the bottom equation of Eqs. (7) a relationship between harmonics of electric field and current $\mathbf{j}_\omega = n_e e^2 \mathbf{E}_\omega / i\omega m_e$ one can write

$$\text{rot }\mathbf{B}_\omega = \frac{i\omega}{c}\left(1 - \frac{\omega_p^2}{\omega^2}\right)\mathbf{E}_\omega . \qquad (12)$$

For infinite plasma in a linear regime $\omega = \omega_p$, which results in that rot $\mathbf{B} = 0$. In an axially symmetric case, this yields that plasma oscillations do not create the azimuthal component of the magnetic field. Note that this conclusion is not correct for a plasma with a boundary where a surface plasma wave is excited, and for high amplitude plasma oscillations. In both cases $\omega < \omega_p$, and magnetic field penetrates into the plasma but exponentially decays with penetration depth.

To find the beam focusing force we assume an azimuthal symmetry of the accelerating field but do not assume that the plasma wave has to have a small amplitude. Rewriting the central equation of Eqs. (6) in the cylindrical coordinates one obtains:

$$\frac{1}{r}\frac{\partial (rE_r)}{\partial r} + \frac{\partial E_z}{\partial z} = -4\pi e(n_e - n_0) . \qquad (13)$$

Taking into account that $E_r|_{r=0} = 0$ and solving Eq. (13) near axis, where one can consider the plasma density and longitudinal electric field being constant, one obtains the plasma electric field near the axis:

$$E_r = -\left(2\pi e(n_e - n_0) + \frac{1}{2}\frac{\partial E_z}{\partial z}\right)r . \qquad (14)$$

Similar using Eq. (11) one gets the magnetic field near the axis:

$$B_\theta = \frac{r}{2c}\left(\frac{\partial E_z}{\partial t} - 4\pi e n_e v_{e_z}\right) , \qquad (15)$$

where $v_{e_z}$ is the longitudinal velocity of plasma electrons near axis. Consequently, the focusing force is:

$$F_r = e\left(E + \frac{[\mathbf{v}_b \mathbf{B}]}{c}\right) = \\ \mp \frac{er}{2}\left(\frac{\partial E_z}{\partial z} + \frac{v_b}{c^2}\frac{\partial E_z}{\partial t} + 4\pi e\left(n_e\left(1 - \frac{v_b v_{e_z}}{c^2}\right) - n_0\right)\right), \qquad (16)$$

where $v_b$ is the bunch velocity, the top and bottom signs should be taken for accelerated positrons and electrons respectively, and the positive longitudinal electric field accelerates positrons. For the wave propagating with velocity of the driving beam $v_d$ one can write that $\mathbf{E}(\mathbf{r}_\perp, z, t) = \mathbf{E}(\mathbf{r}_\perp, z - v_d t)$. It results in:

$$\partial E_z/\partial z + (v_b/c^2)\partial E_z/\partial t = \left(1 - v_b v_d/c^2\right)\partial E_z/\partial z .$$

Consequently, if both the driving and the accelerated beams move with velocities close the speed of light the first two terms in Eq. (16) are cancelled and focusing force is determined by particle density at the axis.

$$F_r = \mp 2\pi e^2 r\left(n_e\left(1 - v_{e_z}/c\right) - n_0\right) . \qquad (17)$$

As one can see, the motion of plasma electrons in the beam direction reduces their contribution to the focusing force. In the bubble regime, when electrons are absent on the axis and the ion density changes inside the bubble can be neglected, the focusing force is constant inside the bubble.

## 3. Emittance Growth due to Multiple Scattering in Plasma

There are two major limitations which require very strong focusing of bunches, which are accelerated in a plasma. The first one is related to multiple scattering of accelerated particles on the plasma ions and electrons (if present); and the second one is related to the BBU instability because of the transverse impedance of the plasma channel. In this section we consider the first limitation.

The normalized emittance growth in a singly-ionized plasma due to multiple Coulomb scattering is given by (in the ultra-relativistic case) [9]:

$$\frac{d\varepsilon_n}{d\gamma} = \frac{2\pi\left(Z(Z+1) + n_e/n_0 - 1\right)r_e^2 n_0 \Lambda_c}{\gamma(d\gamma/ds)}\beta_f(\gamma) , \qquad (18)$$

where $Z$ is the plasma ion charge, $\Lambda_c \approx 18$ is the Coulomb logarithm, and $\beta_f(\gamma)$ is the beta-function of beam focusing.

Below we consider the emittance growth for the most optimistic case of acceleration of electrons in a bubble regime in a fully-ionized hydrogen plasma ($n_e = 0$, $Z=1$). To simplify an estimate we will assume that a harmonic wave with $E = E_0 \cos(\omega t - kz)$ is excited[3]. Here we also assume 100% density modulation resulting in that $E_0$ is determined by Eq. (10), and $k=\omega_p/c$. Then, the acceleration rate is equal to

$$\frac{d\gamma}{ds} \approx \frac{eE_0}{mc^2}\cos\phi_{acc} = \sqrt{4\pi n_0 r_e}\cos\phi_{acc} . \quad (19)$$

where $\phi_{acc}$ is the accelerating phase. Assuming that the beam focusing is supported by the plasma itself one obtains from Eq. (17) the corresponding beta-function:

$$\beta_f(\gamma) = \frac{c\sqrt{2\gamma}}{\omega_p} . \quad (20)$$

---
[3] It is straightforward to extend this calculation for more accurate description of beam acceleration in the bubble regime

Substituting Eqs. (19) and (20) to Eq. (18) and integrating to the final energy with $\gamma$-factor of $\gamma_f$ one obtains the normalized emittance increase due multiple scattering in the course of entire acceleration:

$$\Delta\varepsilon_n = \frac{\sqrt{2\gamma_f}\, r_e \Lambda_c}{\cos\phi_{acc}} . \quad (21)$$

Here, we neglected a contribution to the integral of its lower limit corresponding to the initial beam energy. As one can see from Eq. (21) the emittance growth does not depend on the plasma density.

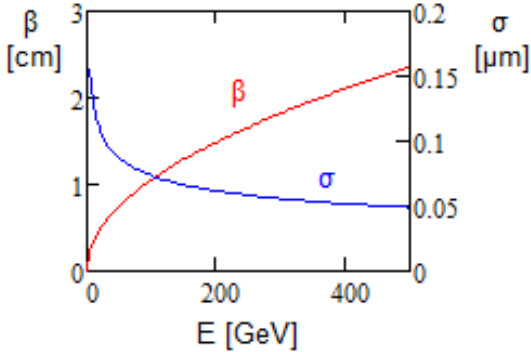

Figure 1: The beta-function and beam size for acceleration in the bubble regime; $n_0 = 10^{17}$ cm$^{-3}$.

For the beam acceleration to 0.5 TeV with the accelerating phase of 60 deg. one obtains $\Delta\varepsilon_n$=0.14 nm. This value is almost three orders of magnitude smaller than the value for the LPA emittance presented in Table 1. However it has been achieved due to extremely strong plasma focusing. For the plasma density of $10^{17}$ cm$^{-3}$ suggested in Ref. [2], one obtains the effective magnetic field gradient of plasma focusing of $3 \cdot 10^4$ T/cm. It is more than 3 orders of magnitude larger than the present state-of-the-art superconducting quadrupoles. This implies that only plasma focusing is capable of suppressing the emittance growth due to multiple scattering on the plasma ions. Figure 1 presents the rms beam size and the beta-function for beam acceleration with LPA parameters. As one can see the rms beam size is well below 1 μm for the major part of acceleration. Finally we need to note that multiple scattering also considered in detail in Sections 6 and 7.

limits the use of heavy plasma ions. In the case of LPA it requires Z≤10 if singly charged ions are used.

## 4. Bunch Deceleration in Uniform Plasma

In this section we consider the deceleration of a short bunch in a uniform plasma. We assume that the transverse beam size is equal to zero. The consequences and the accuracy of this approximation will be considered at the end of this section. To find the decelerating force along the bunch, we will follow the standard procedure presented in Ref. [10]. The approach consists of three steps: (1) finding a collective plasma response at large impact parameters, where the plasma perturbation theory is justified, (2) computing a contribution of small impact parameters, where the plasma perturbation theory is not justified, but the collective response of the plasma can be neglected, and (3) combining two contributions at some medium impact parameter.

To find the collective plasma response we follow Ref. [11]. We consider a point-like charge $q$ moving with velocity $v_b$ along axis $z$ through a plasma. Its charge density can be presented as

$$\rho(\mathbf{r}_\perp, z, t) = q\delta(\mathbf{r}_\perp)\delta(z - v_b t) . \quad (22)$$

To describe a plasma response we add this external charge to the middle equation of Eqs. (7) so that:

$$\text{div}(\mathbf{E}) = 4\pi\left(q\delta(\mathbf{r}_\perp)\delta(z - v_b t) - e\tilde{n}\right) . \quad (23)$$

Excluding $\mathbf{E}$ and $\mathbf{v}$ from the modified Eqs. (7) one obtains:

$$\frac{\partial^2 \tilde{n}}{\partial t^2} + \omega_p^2 \tilde{n} = \frac{4\pi n_0 eq}{m}\delta(\mathbf{r}_\perp)\delta(z - v_b t) . \quad (24)$$

The solution of Eq. (24) is the Green's function of a harmonic oscillator namely,

$$\tilde{n} = \frac{\omega_p q}{ev_b}\sin(\omega_p t - k_p z)\theta(t - z/v_b)\delta(\mathbf{r}_\perp). \quad (25)$$

Here θ(t) is the step function which is 1 or 0 for positive or negative values of its argument, respectively; we also took into account that $\delta(z - v_b t) = \delta(t - z/v_b)/v_b$, and $k_p = \omega_p/v_b$. Transitioning to the bunch frame, one cancels the dependence on time. For $|z - v_b t| \gg r_\perp/\gamma$, a solution of the resulting Poisson equation is straightforward. Coming back to the lab-frame one obtains the resulting scalar potential:

$$\varphi = \begin{cases} -2qk_p K_0(r_\perp k_p)\sin(\omega_p t - k_p z), & -s \gg r_\perp/\gamma, \\ 0, & s \gg r_\perp/\gamma, \end{cases} \quad (26)$$

where $s = z - v_b t$, and $K_0(x)$ is the modified Bessel function, and we also accounted that the negative sign of plasma electrons charge introduces a sign change in Eq. (26). In the ultra-relativistic case $r_\perp/\gamma \to 0$, that results in a replacement of inequalities in Eq. (26) by the θ-function. Setting $v_b = c$ one obtains the corresponding electric field for the ultra-relativistic case ($k_p = \omega_p/c$):

$$E_\parallel = -2qk_p^2 K_0(r_\perp k_p) \cos(\omega_p t - k_p z) \theta(t - z/c),$$
$$E_\perp = -2qk_p^2 K_1(r_\perp k_p) \sin(\omega_p t - k_p z) \theta(t - z/c). \quad (27)$$

Here $K_1(x)$ is the modified Bessel function. The evaluation of rot(**E**) demonstrates that the magnetic field in the trailing wave is equal to zero. It is related to the fact that the magnetic field, excited by the longitudinal component of the plasma current, is compensated by magnetic field excited by changing longitudinal electric field, as can be seen from Eq. (12). As one can see from Eq. (25) the density perturbation is equal to zero everywhere except on the axis. Consequently, everywhere except for the axis, div $\mathbf{v}_e$ = div $\mathbf{E}$ = 0.

The above obtained equations diverge at small $\mathbf{r}_\perp$. It is related to a violation of applicability of the linearized plasma equations. Therefore, a contribution of small impact parameters (small $\mathbf{r}_\perp$) needs to be evaluated separately. First, we consider how the wakefield will be changed if plasma is removed from small impact parameters. To do so, we consider a problem where a charge moves along axis $z$ in a plasma with a round channel of radius $b$ around the particle trajectory. This problem was solved in Ref. [12] (see also Ref. [13][4]). The obtained longitudinal electric field inside the channel is constant across it and is equal to:

$$E_\parallel = -2qk_p^2 K_0(bk_p) \kappa_0(k_p b) \times$$
$$\cos(\Omega(k_p b)\omega_p t - k_p z) \theta(t - z/c), \quad r_\perp < b. \quad (28)$$

The corresponding radial electric and azimuthal magnetic fields are:

$$E_r = H_\theta = qk_p^3 r_\perp K_0(bk_p) \kappa_0(k_p b) \times$$
$$\sin(\Omega_0(k_p b)\omega_p t - k_p z) \theta(t - z/c), \quad r_\perp < b, \quad (29)$$

where the correcting factors are:

$$\Omega_0(x) = \sqrt{\frac{2K_1(x)}{2K_1(x) + xK_0(x)}},$$
$$\kappa_0(x) = \frac{1}{x(K_1(x) + xK_0(x)/2)}. \quad (30)$$

Figure 2 shows plots of these functions. As one can see for $x \ll 1$ both functions are close to 1. Consequently, for $b \ll k_p^{-1}$ we can put them equal to 1. In this case the longitudinal electric field of Eq. (27) and (28) coincide if $r_\perp = b$.

---

[4] It is straightforward to obtain the equations describing bunch interaction with plasma from theory developed in Ref. [13] where the resistive wall impedances of a round channel were found. The replacement of the wall material conductivity $\sigma$ by the plasma conductivity $\sigma_p = -e^2 n_0 / (im\omega)$ addresses the problem.

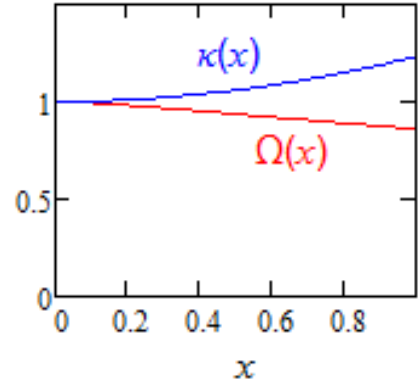

Figure 2: Plots of functions $\Omega_0(x)$ and $\kappa_0(x)$.

Before proceeding to the bunch of a finite length we consider a point-like bunch. The decelerating electric field acting on a charged particle (point-like bunch) traveling through plasma is well known and is derived in a number of textbooks. In particular, it can be found in Ref. 10. For a particle moving with the speed of light, its result can be presented in the following form:

$$E_p = -qk_p^2 \ln\left(\frac{2}{e^{\gamma_E}} \frac{\rho_{max}}{\rho_{min}}\right), \quad (31)$$
$$\rho_{min} = r_e q/e, \quad \rho_{max} = 1/k_p.$$

where $\gamma_E \approx 0.577$ is the Euler's constant ($\exp(\gamma_E) \approx 1.781$). Note that the decelerating electric field immediately behind the particle is twice larger than the decelerating field of Eq. (31). Expanding the Bessel function in Eq.(28) one obtains:

$$E_\parallel(\xi) = -2qk_p^2 \ln\left(\frac{2}{e^{\gamma_E}} \frac{1}{k_p b}\right) \cos(k_p \xi), \quad k_p b \ll 1, \quad (32)$$

where $\xi = ct - z$. Comparing Eqs. (31) and (32) one can see that the results coincide if $b = r_{min}$. This means that the contribution of impact parameters lower than $\rho_{min}$ is greatly reduced relative to the prediction of the linearized plasma theory, i.e. accurate accounting of small impact parameters should remove the divergence at small $r_\perp$ in Eqs. (27)-(29).

For a bunch with finite length, the contribution of small impact parameters depends on the longitudinal particle distribution inside the bunch and, consequently, on the bunch length. As will be seen below, the contribution of small impact parameters also depends on the sign of the charge. If the bunch is much shorter than $k_p^{-1}$, we can neglect collective effects in the plasma and consider the bunch interaction with non-interacting electrons. To further simplify this problem, we consider a bunch with a uniform longitudinal density, the bunch length $L_b$ and the number of particles $N_b$ so that $q = N_b e$. To eliminate the dependence on time, we consider the problem in the bunch frame. In this case the plasma with density $n' = \gamma n_0$ moves along $z'$ axis towards the bunch with the bunch length $L' = \gamma L$, which is much larger than any transverse dimension in the problem. Here and below we use sign ′ to denote values in the bunch frame. In the ultrarelativistic case the equation of motion for a plasma electron in the bunch frame is:

$$\frac{d^2 r_\perp}{ds'^2} = \pm \frac{2e^2 N_b}{mc^2 \gamma^2 L_b} \frac{1}{r_\perp} ,\qquad (33)$$

where signs plus and minus are taken for electron and positron bunches, respectively. Note that this equation is justified even for the case when the transverse motion of plasma electrons is relativistic in the plasma (lab) frame. The only condition of Eq. (33) applicability is that the transverse momentum of plasma electron is smaller than its initial momentum in the bunch frame $p_\perp \ll mc\gamma$. It is always the case for a major part of beam acceleration in a plasma-based collider with possible exception for very small beam energy at the accelerator beginning. Note also that a description of electron motion in the lab frame for a relativistic motion of a plasma electron would require accounting for the electron longitudinal motion excited by the bunch axial magnetic field and a solution of relativistic equations of the electron motion.

First, we consider plasma interaction with an electron bunch. The right-hand side of Eq. (33) does not depend on $s'$ and its integration is straightforward. The first integration yields:

$$\frac{dr_\perp}{ds'} = \sqrt{\frac{4e^2 N_b}{mc^2 \gamma^2 L_b} \ln\left(\frac{r_\perp}{r_0}\right)} ,\qquad (34)$$

where $r_0$ is the impact parameter (electron initial radius). The next integration yields:

$$2r_m \frac{s'}{\gamma L_b} = r_0 \int_1^{r_\perp / r_0} \frac{dx}{\sqrt{\ln(x)}} ,\qquad (35)$$

where

$$r_m = \sqrt{N_b r_e L_b} = \sqrt{\rho_{\min} L_b} .\qquad (36)$$

Let us introduce the function:

$$\mathrm{R}_e(x) = \int_1^x \frac{dx'}{\sqrt{\ln(x')}} = -i\sqrt{\pi}\,\mathrm{erf}\left(i\sqrt{\ln(x)}\right) ,\qquad (37)$$

where $\mathrm{erf}(x) = (2/\sqrt{\pi}) \int_0^x \exp(-t^2) dt$ is the error function. Results of numerical inversion of $\mathrm{R}_e(x)$ can be presented by a fitting formula:

$$\mathrm{R}_e^{-1}(x) \approx 1 + \frac{x^2}{4(1+0.123x)} + \frac{0.055 x^{10/3}}{10^4 + x^2} ,\qquad (38)$$

which has a few percent accuracy in the range $x \in [0, 4000]$ - the range sufficient for most applications. Then, a particle trajectory can be described by the following equation:

$$r_\perp(s', r) = r_0\, \mathrm{R}_e^{-1}\!\left(\frac{2r_m}{r_0} \frac{s'}{\gamma L_b}\right) .\qquad (39)$$

To find the contribution of small impact parameters to the decelerating force, first, we find a scalar potential at the axis. After passing distance $s'$ the plasma electrons contained before collision in a thin cylinder with the radius $r$ and the thickness $dr$ are separated from the ions (those deflection can be neglected) and, consequently, the ions stay at the same radius. The summation of contributions of all radii yields the scalar potential on axis:

$$\varphi(s') = 4\pi\gamma n_0 e \int_0^b \ln(r_\perp(s', r)/r) r\, dr .\qquad (40)$$

Consequently, the longitudinal electric field excited at the axis is:

$$E_\parallel(s') = -\frac{d\varphi}{ds} = -4\pi\gamma n_0 e \int_0^b \frac{dr_\perp}{ds'} \frac{r}{r_\perp(s', r)} dr .\qquad (41)$$

Substituting $r_\perp(s', r)$ from Eqs. (39), transiting to a dimensionless variable in the integration and, finally, returning to the lab frame we obtain the contribution to the longitudinal electric field at the axis produced by collisions with small impact parameters:

$$E_\parallel(\xi) = 2e N_b k_p^2 \frac{\xi}{L_b} \Phi_e\!\left(\frac{L_b b}{2 r_m \xi}\right), \quad \xi = s'/\gamma ,\qquad (42)$$

where the sign is changed because $s$ and $s'$ are directed in opposite directions, and

$$\Phi_e(x) = 2\int_0^x \frac{\mathrm{D}_e(1/x')}{\mathrm{R}_e^{-1}(1/x')} dx', \quad \mathrm{D}_e(x) = \frac{d}{dx}\mathrm{R}_e^{-1}(x) .\qquad (43)$$

The function $\Phi(x)$ was calculated by a numerical integration, whose results can be approximated by the following equation:

$$\Phi_e(x) \approx \ln\!\left(1 + \frac{2.93 x^2}{1 + 2x}\right) .\qquad (44)$$

To find the electric field coming from large impact parameters one needs to sum contributions for all particles in the bunch. Integrating Eq. (32) so that only contributions for upstream particles is accounted one obtains:

$$E_\parallel(\xi) = 2e N_b k_p^2 \frac{\xi}{L_b} \ln\!\left(\frac{2}{e^{\gamma_E}} \frac{1}{k_p b}\right), \quad \begin{cases} \xi k_p \ll 1, \\ b k_p \ll 1. \end{cases}\qquad (45)$$

Leaving only the leading term in Eq. (42) and summing it with Eq. (45) one finally obtains, in logarithmic approximation, the total decelerating electric field along the electron bunch with uniform longitudinal distribution:

$$E_\parallel(\xi) \approx 2e N_b k_p^2 \frac{\xi}{L_b} \ln\!\left(\frac{2.93}{e^{\gamma_E}} \frac{1}{r_m k_p} \frac{L_b}{\xi}\right), \quad \begin{cases} \xi k_p \ll 1, \\ r_m k_p \ll 1. \end{cases}\qquad (46)$$

As one can see a dependence on $b$ disappeared. That supports an applicability of the method where collisions at small and large impact parameters are accounted separately. Figure 3 shows the electric field along the electron bunch for different values of $r_m k_p$.

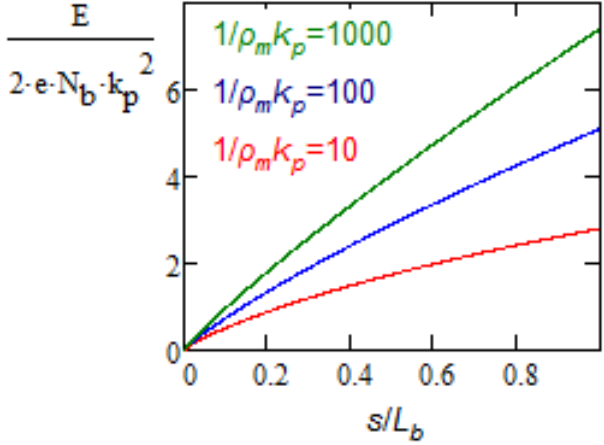

Figure 3: Dependence of decelerating field along bunch for different values of $\rho_m k_p$.

Now, we consider plasma interaction with the positron bunch, where we will follow the procedure used above for the electron bunch. In this case the solution of Eq. (33) is:

$$2r_m \frac{s'}{\gamma L_b} = r_0 \int_1^{r_\perp / r_0} \frac{dx}{\sqrt{-\ln(x)}} \ . \quad (47)$$

Introducing function,

$$R_p(x) = -\int_1^x \frac{dx'}{\sqrt{-\ln(x')}} = \sqrt{\pi}\, \mathrm{erf}\left(\sqrt{-\ln(x)}\right), \quad (48)$$

we obtain the particle trajectory:

$$r_\perp(s', r) = r_0 R_p^{-1}\left(\frac{2r_m}{r_0} \frac{s'}{\gamma L_b}\right), \quad (49)$$

where $R_p^{-1}(x)$ is the inverse of function $R_p(x)$, and it can be approximated as following:

$$R_p^{-1}(x) \approx \left(\left(\frac{9}{8} - \frac{1}{3\pi}\right)\cos\left(\frac{\sqrt{\pi}}{2}x\right) + \left(\frac{1}{3\pi} - \frac{1}{8}\right)\cos\left(\frac{3\sqrt{\pi}}{2}x\right)\right)^{\frac{3}{4}}. \quad (50)$$

The same as for electrons, Eq. (42) describes the electric field at the axis with function $\Phi(x)$ replaced by function:

$$\Phi_p(x) = \int_0^x F\left(\frac{1}{x'}\right) dx', \quad (51)$$

$$F(x) = 2\frac{D_p(x)}{R_p^{-1}(x)}, \quad D_p(x) = \frac{d}{dx} R_p^{-1}(x).$$

To eliminate the divergence in the integrand we replace variable so that $x \to 1/x$, split area of integration at pieces equal to a half period, and group addends of opposite signs together:

$$\Phi_p(x) = \int_{1/x}^\infty \frac{F(x')}{x'^2} dx' = \int_{1/x}^{\sqrt{\pi}/2} \frac{F(x')}{x'^2} dx' + \int_{\sqrt{\pi}/2}^{2\sqrt{\pi}} \frac{F(x')}{x'^2} dx'$$

$$+ \int_0^{\sqrt{\pi}/2} \left(\frac{F(\sqrt{\pi} + x')}{(\sqrt{\pi} + x')^2} + \frac{F(\sqrt{\pi} - x')}{(\sqrt{\pi} - x')^2}\right) dx' \quad (52)$$

$$+ \sum_{k=1}^\infty \int_0^{\sqrt{\pi}} \left(\frac{F(\sqrt{\pi} + x')}{((1+2k)\sqrt{\pi} + x')^2} + \frac{F(\sqrt{\pi} - x')}{((1+2k)\sqrt{\pi} - x')^2}\right) dx',$$

$$1/x \leq \sqrt{\pi}/2.$$

Here we used periodicity of $F(x)$ with period $2\sqrt{\pi}$ and that $F((1+2k)\sqrt{\pi} + x) = -F((1+2k)\sqrt{\pi} - x)$. Results of numerical integration can be approximated as

$$\Phi_p(x) \approx -\ln\left(1 + \frac{3.02 x^3}{x^2 + 1.05}\right). \quad (53)$$

Finally, we obtain the contribution of small impact parameters,

$$E_\parallel(\xi) = 2 e N_b k_p^2 \frac{\xi}{L_b} \Phi_p\left(\frac{L_b b}{2 r_m \xi}\right), \quad (54)$$

and the total decelerating field for positron bunch in the logarithmic approximation:

$$E_\parallel(\xi) \approx -2 e N_b k_p^2 \frac{\xi}{L_b} \ln\left(\frac{3.02}{e^{\gamma_E}} \frac{1}{r_m k_p} \frac{L_b}{\xi}\right), \quad \begin{cases} \xi k_p \ll 1, \\ r_m k_p \ll 1. \end{cases} \quad (55)$$

Same as for electrons, the major contribution to the decelerating force comes from plasma electrons with impact parameters larger than $r_m$.

We need to note that an assumption that an interaction of plasma electrons can be neglected is not quite accurate in the case of positron beam. Plasma electrons are attracted to the axis and create very large charge density in its close vicinity. For electrons with sufficiently small impact parameters it results in their repulsion from the axis. However, it has little effect on the decelerating force due to overall small contribution of small impact parameters, $r < r_m$. It is important to note that large density of plasma electrons destroys linearity of plasma focusing discussed in Section 2 and greatly amplifies multiple scattering discussed in Section 3. Each of this effects makes impossible an acceleration of collider quality positron bunch in plasma-based accelerator.

The decelerating forces for positrons and electrons are quite similar for small bunch population when $k_p r_m$ is small. However with intensity increase, the decelerating force for positron bunch is up to ~45% larger as shown by ratio $\Phi_p\left(1/e^{\gamma_E} r_m k_p\right) / \Phi_e\left(1/e^{\gamma_E} r_m k_p\right)$ in Figure 4.

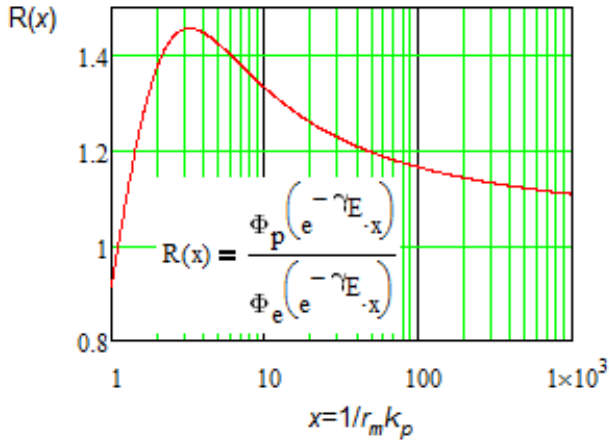

Figure 4: The ratio of deceleration forces for positron and electron bunches at the bunch tail as function of $r_m k_p$.

The $r_m$ represents the impact parameter where the radial displacement of a plasma electron during its passage through the bunch is about its initial value. It is demonstrated in Figure 5 presenting a plasma electron trajectory in the course of collision with electron and positron bunches. For both cases, the major contribution to the decelerating field comes from large impact parameters, $r > r_m$. Therefore if the transverse bunch size is significantly smaller than $r_m$ its value does not affect the decelerating field. We would like to stress that for a bunch brightness required for a collider this condition is always satisfied; and one can consider that, in finding the deceleration force, a bunch has zero transverse size.

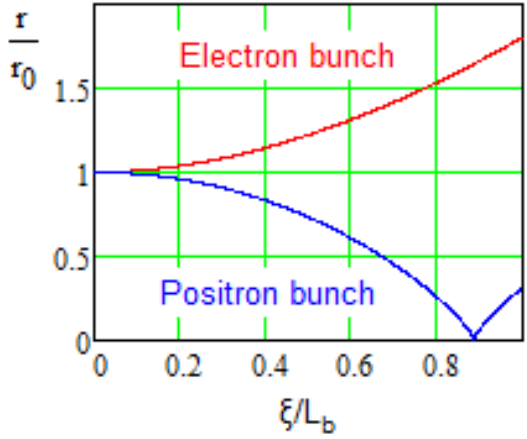

Figure 5: The trajectory of a plasma electron for collisions with electron and proton bunches for the impact parameter equal to $r_m$, i.e. $r_0 = r_m$.

As one can see from Eqs. (46) and (55) there is a logarithmic dependence of decelerating force on $\xi$. That means that a contribution to the decelerating force left behind by a specific particle depends on its longitudinal position in the bunch. Such dependence is related to plasma density perturbation produced by previous particles. As one can see from Figure 3 it becomes less significant with reducing $r_m k_p \ll 1$. Note also that this effect is much less significant for a bunch with Gaussian distribution.

The decelerating forces of Eqs. (46) and (55) are justified for $r_m k_p \ll 1$. As one can see they change sign when the argument in logarithm becomes equal to 1. This happens because the maximum impact parameter, $k_p^{-1}$, becomes smaller than the minimum impact parameter $r_m$. An addition of $r_m$ to the maximum impact parameter, or in other words an addition of 1 to the arguments in logarithms, corrects this problem and yields reasonably good estimates of the decelerating field even for the case of bubble regime when $r_m k_p \geq 1$. A comparison of numerical simulations presented in Ref. [14] with calculations in the present paper show reasonable coincidence. In the case $r_m k_p \gg 1$, the logarithm can be approximated by a linear function. Then, using Eq. (46) one obtains the decelerating force for electron beam:

$$E_\parallel(\xi) \approx 3.2 e k_p \sqrt{\frac{N_b}{r_e L_b}}, \quad \frac{2 \cdot 2.93}{e^{\gamma_E}} \to 3.2, \quad \begin{cases} \xi k_p \ll 1, \\ r_m k_p \gg 1. \end{cases} \quad (56)$$

As one can see in the strong bubble regime the decelerating field grows proportionally to $\sqrt{N_b}$. We will see this dependence in Section 7 where the bunch acceleration in the strong bubble regime is considered.

The above calculations describe the decelerating force for the case of $L_b k_p \ll 1$. In linear approximation a plasma response results in an oscillation of this field with distance from the bunch at the space frequency of $k_p$. In most of practical cases such approximation is not quite valid. Even a small charge electron bunch moving in a plasma pushes plasma electrons out its way thus creating a cavity, void of plasma electrons.

An evaluation of radial size and length of the cavity created by a point-like bunch is considered in Ref. [15] for the case when the motion of plasma electrons is non-relativistic, $N_b k_p^3 / (4\pi n_0) \ll 1$. The paper concludes that the length of the cavity does not depend on the number of particles in the bunch and is equal to $3.8/k_p$, and its radial size is

$$r_{cav} \approx 2.82 \sqrt{N_b} \sqrt[4]{\frac{r_e}{4\pi n_0}} = 2.82 \frac{r_m}{\sqrt{k_p L_b}}. \quad (57)$$

Although this result was obtained for zero bunch length it is actually justified for $L_b k_p \leq 0.4$. In the range $0 \leq L_b k_p \leq 0.4$ the cavity radius is decreasing approximately linearly with the bunch length and achieves ~80% of its zero-length value at $L_b k_p = 0.4$. From the second half of Eq. (57) one can see that $r_{cav} > r_m$ for $L_b < k_p^{-1}$.

The minimum impact parameter, $r_m$, depends on the longitudinal density distribution in the bunch. To obtain the decelerating force for the Gaussian distribution we followed the described above procedure. Finding particle trajectories in the first order perturbation theory, substituting them to Eq. (41), performing integration over impact parameter, and adding the contribution of large impact parameters one obtains the longitudinal field with logarithmic accuracy:

$$E_\parallel(\xi) \approx -eN_b k_p^2 \left(1 + \mathrm{erf}(\hat{\xi})\right) \times \ln\left(\frac{2\sqrt{2}/(e^{\gamma_E} r_{mg} k_p)}{\sqrt{\sqrt{\pi}\hat{\xi}\left(1+\mathrm{erf}(\hat{\xi})\right) + \exp(-\hat{\xi}^2)}}\right), \quad \hat{\xi} = \frac{\xi}{\sqrt{2}\sigma_s}, \quad (58)$$

where $r_{mg} = \sqrt{2\sqrt{2/\pi}\, N_b r_e \sigma_s}$ is the minimum impact parameter for beam with Gaussian distribution, $\sigma_s$ is the bunch rms length, and we assume that the bunch center is at $\xi = 0$. One can see that similar to the rectangular particle distribution considered above there is a logarithmic dependence of single particle response on its longitudinal coordinate.

## 5. Longitudinal and Transverse Wakes in Plasma

To create an accelerating field in plasma one has to excite it by a short electron bunch or a short pulse of laser radiation. In the context of this section the method of excitation is not important. The excitation creates a cavity moving with the velocity close to the speed of light. The cavity's electric field accelerates and focuses particles of the accelerated (witness) bunch. The longitudinal electric field experiences a half period oscillation along the cavity changing from deceleration of electrons in its first half to the acceleration in the second half so that the maximum acceleration is achieved at the cavity end. In this section we consider an interaction of a bright electron bunch with a plasma surrounding the cavity.

First, we ignore that the cavity walls are moving and replace them by a plasma channel considered at the beginning of the previous section. Then it is straight forward to obtain the longitudinal wake function from Eq. (28). At small distances ($\xi \ll k_p^{-1}$) it is constant and is equal to:

$$W_\parallel \equiv -\frac{E_\parallel(0^+)}{q} = \frac{2k_p^2 K_0(R)}{R(K_1(R) + RK_0(R)/2)}, \quad 0 < \xi \ll k_p^{-1}. \quad (59)$$

where $R = k_p \tilde{b}$, and $\tilde{b} = \max(b, r_m)$ is the maximum between the channel radius and the minimum impact parameter introduced in Eq. (35). Figure 6 presents the plot of this function and its asymptotes:

$$W_\parallel \approx 2k_p^2 \begin{cases} \ln(1 + 2/e^{\gamma_E} k_p \tilde{b}), & 0 \le \tilde{b} k_p \le 0.3, \\ 2/(1 + k_p \tilde{b})^2, & \tilde{b} k_p \ge 1. \end{cases} \quad (60)$$

For sufficiently large $R \ge 10$ the dependence of the wake on $k_p$ becomes insignificant:

$$W_\parallel \approx 4/b^2. \quad (61)$$

In the case of the bubble regime the electron density near bubble boundary is significantly higher than in the considered here plasma channel. It increases the local $k_p$ and extends the applicability of Eq. (61) to smaller values of $R$. The same result is obtained in Ref. [16] where the longitudinal wake is obtained for the large-size bubble where the motion of plasma electrons is relativistic.

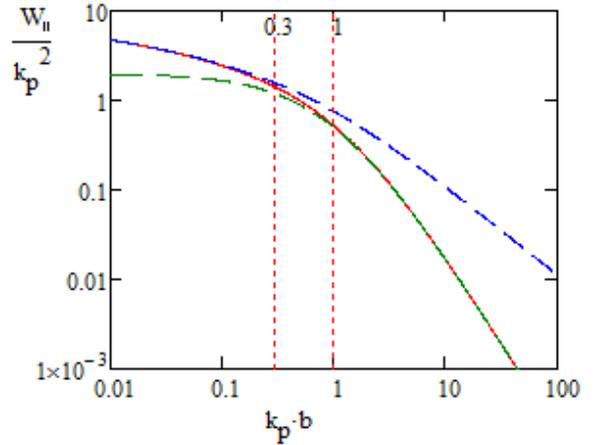

Figure 6: Dependence of longitudinal wake function on parameter $R = \tilde{b} k_p$ (red line); blue and green line represent the asymptotes at low and high values of $R$, respectively.

To find the transverse wake we will split the problem into two steps - separately considering collisions with small and large impact parameters following the recipe of the previous section. However, in the evaluation of small impact parameters contribution we will keep only linear terms in the plasma response so that to keep the obtained result in a simple form. If required, it is straightforward to include non-linear terms using the perturbation theory.

In the evaluation of small impact parameters contribution, similar to the previous section, we consider collisions in the bunch frame and will neglect reaction of plasma on the motion of plasma electrons. In this case a plasma with density distribution $n(r)$ encounters a point-like charge $q$ displaced from the axis by distance $a$ in the horizontal plane. The electrons of plasma are scattered on the charge and transverse motion of ions is neglected. In the first order of perturbation theory the scattering angle of electron located at radius $r \gg a$ is:

$$\theta' \approx \frac{2eq}{mc^2\gamma r}\left(1 + \frac{a}{r}\cos\phi\right), \quad (62)$$

where $\phi$ is the angle between directions of $a$ and $r$, and we take into account that $a$ and $r$ are Lorentz-invariant. The first and second terms in Eq. (62) are responsible for creating longitudinal and transverse wakes, respectively. To find the transverse electric field excited near axis by plasma electrons we split incoming plasma beam into thin cylinders with thickness $dr$. Taking into account that the ions in the cylinder are not scattered we obtain the transverse electric field coming from one cylinder:

$$\Delta E'_\perp = 2\pi e\gamma \frac{dn}{dr} s'\theta'_1 \Delta r = 2\pi \gamma e \frac{dn}{dr} s' \Delta r \left(\frac{2eqa}{mc^2\gamma r^2}\right), \quad (63)$$

where $\theta'_1 = 2eqa/(mc^2\gamma r^2)$ is the maximum of second order deflecting angle. Integrating over impact parameters and returning to the lab frame we obtain the transverse impedance:

$$W_\perp = \frac{4\pi e^2 s}{mc^2} \int \frac{1}{r^2} \frac{dn}{dr} dr, \quad (64)$$

where we took into account that $W_\perp = \gamma \Delta E'_\perp /(qa)$ and $\xi = s'/\gamma$. In the case of a plasma channel of radius $b$ with a rigid boundary $dn/dr = n_0 \delta(r-b)$ and the integral evaluation is straightforward:

$$W_\perp = \frac{k_p^2 \xi}{b^2}, \quad \begin{array}{l} k_p \xi \ll 1, \\ k_p b \ll 1. \end{array} \quad (65)$$

This result is justified only for the case when the reaction of plasma can be neglected, i.e. $k_p \xi \ll 1, k_p b \ll 1$.

The transverse impedance of a plasma channel was found in Ref. [12]. The result can be presented in the following form:

$$W_\perp = \frac{k_p^2}{b^2} \kappa_1(k_p b) \frac{\sin\left(\Omega_1(k_p b) k_p \xi /\sqrt{2}\right)}{\Omega_1(k_p b) k_p /\sqrt{2}},$$

$$\kappa_1(x) = \frac{8K_1(x)}{R(4K_2(x) + RK_1(x))}, \quad (66)$$

$$\Omega_1(x) = \sqrt{\frac{4K_2(x)}{4K_2(x) + RK_1(x)}}.$$

Here $K_2(x)$ is the modified Bessel function. Plots of the functions $\kappa_1(x)$ and $\Omega_1(x)$ and their asymptotes at large $x$ are shown in Figure 7. One can see that Eqs. (65) and (66) are identical in the parameter range where Eq. (65) is applicable, $k_p \xi \ll 1, k_p b \ll 1$. This proves that for a plasma channel of sufficiently small radius, $b \ll k_p^{-1}$, one can separately consider collisions at small and large impact parameters and then combine the obtained results. In particular, it creates a straightforward procedure to make an accurate evaluation of the transverse impedance for more complicated cases like if the plasma boundary is not rigid or the motion of plasma electrons is excited at sufficiently small radii. However such a procedure only is applicable for channels of sufficiently small radius, $r \ll k_p^{-1}$, and intensity $r_m \ll k_p^{-1}$.

Finally we would like to note that for a large bubble size, $b \gg k_p^{-1}$, the transverse wake is determined by a collective plasma response and is equal to

$$W_\perp \approx 8\xi/b^4, \quad bk_p \gg 1, \quad \xi k_p \ll 1. \quad (67)$$

Similarly to the longitudinal wake, it does not depend on plasma parameters. It is important to note that for $b \gg k_p^{-1}$ both the longitudinal and transverse wakes are strongly suppressed relative to the predictions of a binary collisions model, and that this model is not applicable for a large size bubble (plasma channel). For $b \gg k_p^{-1}$, all currents in plasma are concentrated in a thin layer near channel boundary, $b$. It brings us to the well-known relationship between impedances, $W_\perp \approx 2W_\parallel \xi/b^2$, which follows from the Panofsky-Wenzel theorem.

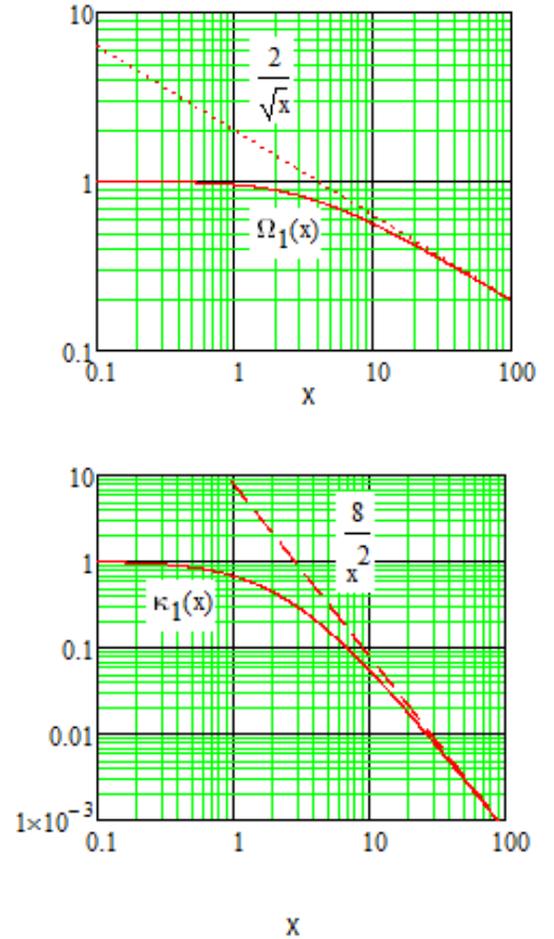

Figure 7: Plots of functions $\Omega_1(x)$ (top) and $\kappa_1(x)$ (bottom).

## 6. Strong Bubble Regime

As one could see from the previous chapters, acceleration of a collider quality electron bunch in the quasi-linear regime and acceleration of a bright positron bunch are not presently feasible. Therefore in this chapter we will focus on the limitations of electron bunch acceleration in the strong bubble regime ($b \gg k_p^{-1}$). In this case, an accurate quantitative description can only be obtained with numerical simulations. However, a qualitative description based on the above described approaches and Ref. [14] delivers reasonably accurate results and, what is even more important, allows one to look into a relationship of different limitations and shows possible ways for optimization of plasma acceleration.

As will be seen below, the main parameter determining the beam transverse stability is the ratio of transverse and longitudinal impedances. It decreases fast with increasing size of the plasma channel, $b$, and thus, it is preferable for beam acceleration. Therefore in this section we consider beam acceleration in the strong bubble regime, when the motion of plasma electrons is relativistic and, consequently, the maximum transverse size of the bubble, $R_b$, is much larger than $k_p^{-1}$. The corresponding equation, which describes the dependence of the bubble radius, $r_b$, on the longitudinal coordinate was developed in Ref. [14]:

$$r_b \frac{d^2 r_b}{d\xi^2} + 2\frac{dr_b}{d\xi}^2 + 1 = \frac{2}{\pi n_e r_b^2}\frac{dN_b}{d\xi}, \quad (68)$$

where $dN_b/d\xi$ is the longitudinal particle density. Below we will call this equation the Lu equation.

In the absence of particles the right-hand side is equal to zero. An introduction of variable $\hat{p} = dr_b/d\xi$ allows one to rewrite Eq. (68) in the following form:

$$r_b \hat{p} \frac{d\hat{p}}{dr_b} + 2\hat{p}^2 + 1 = 0 . \quad (69)$$

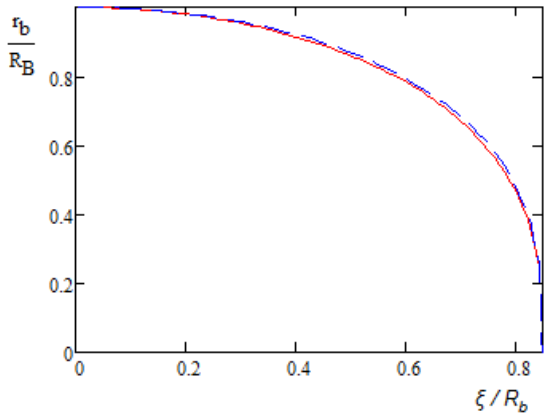

Figure 8: The half bubble profile of Eq. (70) and its fitting by Eq. (72) are presented by solid and dashed lines, respectively.

Its integration yields:

$$\frac{dr_b}{d\xi} \equiv \hat{p} = \pm\sqrt{\frac{1}{2}\left(\frac{R_b^4}{r_b^4} - 1\right)}, \quad (70)$$

where we took into account that $dr_b/d\xi = 0$ at $r_b = R_b$, and signs "+" and "–" are used for the upstream and downstream halves of the bubble. Consequently, the half-bubble length is equal to:

$$\xi_b = \sqrt{2}\int_0^{R_b}\left(\frac{R_b^4}{r^4} - 1\right)^{-1/2} dr = \frac{1}{\sqrt{\pi}}\Gamma\left(\frac{3}{4}\right)^2 \approx 0.847 R_b. \quad (71)$$

The shape of the bubble and its fitting with following approximation,

$$r_b \approx R_b \sqrt[3]{1 - \left(\frac{\xi}{\xi_b}\right)^2}, \quad (72)$$

are shown in Figure 8. Here we assume that the bubble center is located at $\xi = 0$.

Ref. [14] also suggests the equation for the evaluation of the longitudinal electric field in the bubble:

$$E_\| = 2\pi e n_0 r_b \frac{dr_b}{d\xi}. \quad (73)$$

It is straightforward to obtain this equation in the bunch frame (see details in Section 4) assuming that all plasma elections pushed out by the beam are located in the thin layer near the bubble boundary. Note that the focusing part of the electromagnetic field is determined by Eq. (17) with $n_e=0$. Substituting $dr_b/dx$ of Eq. (70) one obtains the field in the bubble:

$$E_\| = \pm\pi e n_0 r_b \sqrt{2\left(\frac{R_b^4}{r_b^4} - 1\right)} \approx \\ \pm\pi e n_0 R_b \sqrt{2\left[\left(1 - \left(\frac{\xi}{\xi_b}\right)^2\right)^{-\frac{2}{3}} - \left(1 - \left(\frac{\xi}{\xi_b}\right)^2\right)^{\frac{2}{3}}\right]}, \quad (74)$$

where signs ± are related to the first and second half of the bubble, and we used Eq. (72) to obtain the second half of Eq. (74). Figure 9 presents a numerically computed dependence of this electric field on coordinate within bubble. Eq. (68) yields that $d^2r_b/d\xi^2 = -1/R_b$ in the bubble center. Using this result in the top part of Eq. (74) one obtains the field gradient in the bubble center:

$$\frac{dE_\|}{d\xi} = -2\pi e n_0 . \quad (75)$$

As one can see it does not depend on the bubble radius.

To obtain the longitudinal wake at short distances we substitute the particle density as $q\delta(\xi - \xi_1)$ to Eq. (68), integrate in close vicinity of $\xi_1$ and substitute the obtained result to Eq. (73). The longitudinal wake obtained this way coincides with the wake of Eq. (61) with replacement $b$ by $r_b$, i. e. $W_\| = 1/r_b^2$. To obtain the dependence of the wake on the coordinate within bubble one needs to integrate Eq. (70). Figure 10 shows wakes numerically calculated for "leading" particles located at different coordinates $\xi_1$. The results can be roughly approximated as:

$$W_\|(\xi, \xi_1) \approx \frac{4}{r_b(\xi)^2}\theta(\xi - \xi_1) . \quad (76)$$

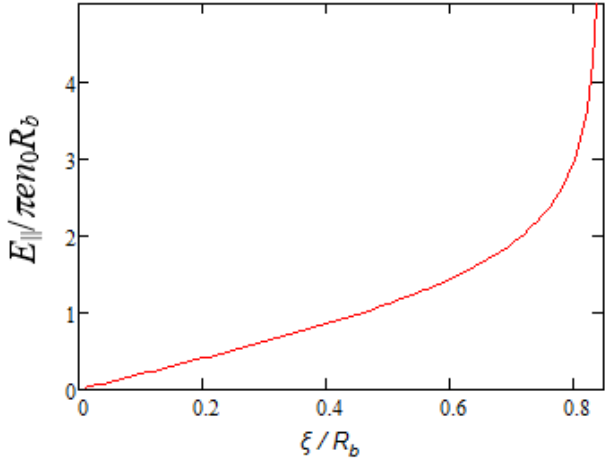

Figure 9: Dependence of dimensionless longitudinal electric field, $E_\parallel/\pi e n_0 R_b$, on the longitudinal coordinate for the bubble trailing half; $\xi = 0$ corresponds to the bubble center.

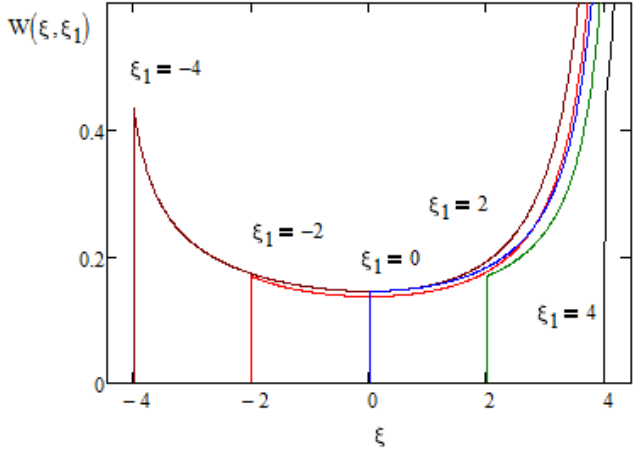

Figure 10: Dependence of longitudinal wake on coordinate within bubble for different points of excitation, $\xi_1$; $R_b = 5$, the bubble center is at $\xi = 0$.

As one can see, the wake does not change its sign to the end of the bubble and diverges at its end. It is completely different from the weak bubble case when the motion of plasma electrons is non-relativistic and the bubble length does not depend on its radius. In the strong bubble case the duration of the bubble is increasing with additional excitation. Combined with independence of the maximum gradient on the excitation strength, it makes the wake being positive for the entire length of the unperturbed bubble. Also note that Eq. (68) is non-linear and, consequently, an introduction of the longitudinal wake is justified for small perturbations only.

## 7. Beam Acceleration in the Strong Bubble Regime

Now we consider a plasma excitation and beam acceleration in the strong bubble regime. First, we require that all particles in the driving bunch would be decelerated with the same rate $E_d$. In this case an integration of Eq. (73) binds up the bubble radius and electric field:

$$E_d \xi = \pi n_0 e \left(r_b^2 - r_{b0}^2\right), \qquad (77)$$

where $r_{b0}$ is the initial bubble radius. Taking into account that at the bunch head, $\xi=0$ and $r_{b0}=0$, expressing $r_b$ from the obtained equation, substituting it to Eq. (68) and performing simple calculations one obtains the longitudinal particle distribution,

$$\frac{dN_b}{d\xi} = \frac{E_d}{8e}\left(\frac{E_d}{\pi n_0 e} + 4\xi\right), \qquad (78)$$

and the total number of particles in the bunch,

$$N_b = \frac{E_d}{8e} L_b^2 \left(\frac{E_d}{\pi n_0 e L_b} + 2\right). \qquad (79)$$

Eq. (77) results in that

$$L_b = \left.\frac{r_b}{2\, dr_b/d\xi}\right|_{r_b=r_{bd}}, \quad \text{and} \quad E_d L_b = \pi n_0 e r_{bd}^2,$$

where $r_{bd}$ is the bubble radius at the end of driving bunch. Using these equations one can rewrite Eq. (79) in the following form:

$$N_b = \frac{\pi n_0 r_{bd}^3}{8}\left(2\frac{dr_{bd}}{d\xi} + \frac{1}{dr_{bd}/d\xi}\right), \qquad (80)$$

which results in the total power transferred from the beam to the plasma:

$$P = e N_b E_d c = \frac{\pi^2}{4} e^2 n_0^2 c R_b^4. \qquad (81)$$

Here we expressed the electric field of through $r_{bd}$ and $dr_{bd}/dx$ using Eq. (73), and $dr_{bd}/dx$ through $R_b$ and $r_b$ using Eq. (70). As one can see the power is uniquely determined by the maximum bubble radius.

One can see from Eq. (68) that if density distributions in the driving and accelerated bunches are mirror symmetric relative to the bunch center, the accelerating bunch is accelerated with the same rate as the deceleration of the driving bunch and 100% of its energy will be transferred to the accelerated (witness) bunch. Actually the ratio of the acceleration to the deceleration forces can be arbitrary as long as the particle distributions and bunch lengths are determined by Eqs. (78) and (79). In this case, if the accelerated bunch extends to the cavity end, the 100% energy transfer efficiency is achieved.

It is important to note that Eq. (68) is approximate and one has to be cautious about the accuracy and meaning of the obtained results. There is a striking difference between the statement in this section that the decelerating force is constant along bunch with a trapezoidal distribution and conclusions of Section 4 (see Eq. (46)) that the decelerating force is zero for the very head of a bunch. This difference originates from the poor description of the plasma reaction by the Lu equation for small $r_b$, for which the longitudinal wake diverges as $1/r_b^2$. Note that in a more accurate model the divergence will be limited by minimum impact parameter. Therefore, Eq. (68) works well only for $L_b \geq k_p^{-1}$. In this case, the absence of deceleration for the bunch head becomes insignificant for the bubble formation and particle acceleration for the rest of the bunch. We also need to stress that Eq. (68) conserves energy exactly. In a real plasma, the accuracy of this statements is improving with increase of $R_b$ however even for large $R_b$ ($R_b \gg k_p^{-1}$) it is not perfectly accurate. Numerical simulations carried out in Ref. [17] show the efficiency of acceleration being ~90% for $k_p R_b \approx 5$.

For a given number of particles the decelerating electric field can be obtained from Eq. (79):

$$E_d = \sqrt{(\pi e n_0 L_b)^2 + \frac{8\pi e^2 n_0 N_b}{L_b}} - \pi e n_0 L_b. \quad (82)$$

For $N_b \geq n_0 L_b^3$ the second addend in the square root dominates and the decelerating field becomes growing proportionally to the square root of bunch population (see Eq. (56)). With the use of above derived relationships, one, after comparatively simple calculations, can express the bubble radius through the bunch length and population:

$$R_b = \frac{L_b}{\sqrt[4]{2}} \sqrt[4]{\frac{8 N_b}{\pi n_0 L_b^3} \left( \sqrt{\frac{8 N_b}{\pi n_0 L_b^3} + 1} - 1 \right)}. \quad (83)$$

Here we assume that the particle distribution is chosen so that to keep constant the decelerating field along the entire bunch. One can see that $R_b \to \infty$ for $L_b \to 0$ which is obviously an incorrect result related to a violation of Eq. (68) applicability ($L_b \geq k_p^{-1}$).

Now we consider limitations on the acceleration of accelerated (witness) bunch. High efficiency of acceleration requires this bunch to extend to the very end of the bubble but it creates a problem with BBU instability because the transverse impedance diverges as $1/r_b^4$ and becomes very large at the cavity end. Therefore the tail of accelerated bunch should be located sufficiently far from the bubble end.

Similarly, let us assume that the particle density is chosen so that all particles are accelerated at the same rate. In this case similar to Eq. (78) the particle distribution is trapezoidal but with particle density linearly decreasing to the bunch tail. Expressing coordinates of bunch head and tail through the bubble radii at their locations with usage of Eq. (77) one obtains the total number of particles in the accelerated bunch:

$$N_a = \frac{E_a}{8e} \left( r_{b2}^2 - r_{b1}^2 + 2 \left( \frac{\pi n_0 e}{E_a} \right)^2 (r_{b2}^4 - r_{b1}^4) \right). \quad (84)$$

Here $E_a$ is the accelerating field, $r_{b2}$ and $r_{b1}$ are the bubble radii in the locations of bunch head and tail, respectively. Then, the power transferred to the accelerated bunch is:

$$P_a = \frac{\pi^2 e^2 n_0^2 c}{4} (r_{b2}^2 - r_{b1}^2) \left( \frac{R_b^4}{r_{b2}^2} + r_{b1}^2 \right). \quad (85)$$

To obtain this equation we expressed the accelerating field in Eq. (84) through the bubble radius using Eq. (74). Consequently, the efficiency of power transfer from the driving to accelerated bunch is:

$$\eta_P = \frac{P_a}{P} = \frac{r_{b2}^2 - r_{b1}^2}{R_b^2} \left( \frac{R_b^2}{r_{b2}^2} + \frac{r_{b1}^2}{R_b^2} \right). \quad (86)$$

Figure 11 shows an example, illustrating the bubble shape and the particle distributions of the driving and accelerated bunches for the power transfer efficiency of 50% and the transformer ratio of 2. For $n_0 = 10^{17}$ cm$^{-3}$ the driving bunch parameters are chosen to be $R_b k_p = 5$, $L_b k_p = 2.5$ yielding the decelerating field of $E_d = 50$ GV/m and $N_b = 3.55 \cdot 10^{10}$. The accelerated bunch parameters are: $r_{b2} = 0.518 R_b$, $r_{b1} = 0.373 R_b$, $E_a = 100$ GV/m, $N_a = 8.86 \cdot 10^9$.

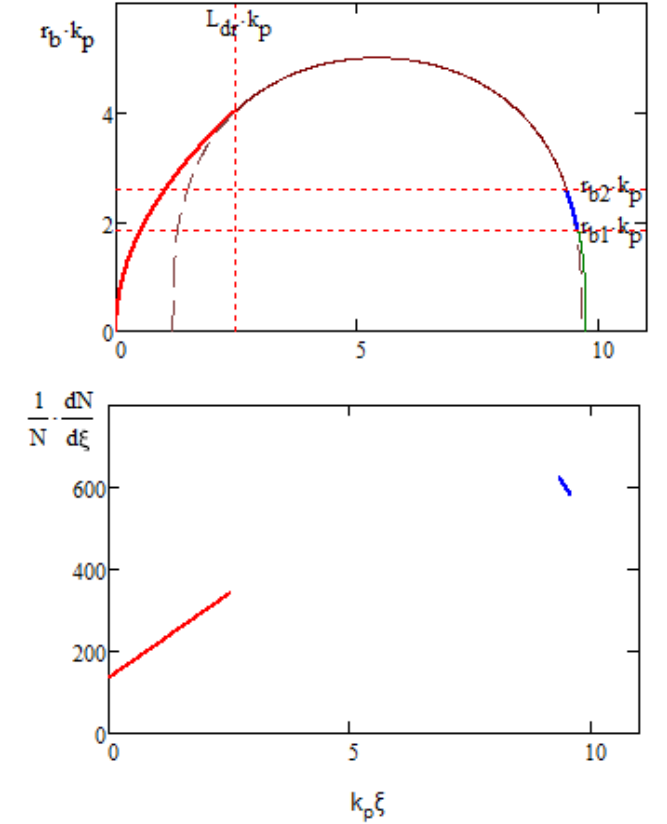

Figure 11: Bubble shape (top) and particle distributions for driving and accelerated bunches (bottom). Red and blue lines are related to the driving and accelerated bunches, respectively. Dashed brown line shows the bubble shape in the absence of particles inside the bubble. Accelerating and decelerating fields are constant: $E_d/E_0 = 1.64$, $E_a/E_0 = 3.28$.

# 8. Beam-breakup Instability in the Strong Bubble Regime

In this chapter we consider a beam-breakup (BBU) instability for an accelerated (witness) bunch. The main parameter determining the development of BBU is the ratio of the bunch deflection force to the focusing force. Focusing is completely dominated by plasma and is described by Eq. (17) where in the bubble regime $n_e=0$. This yields:

$$F_r = -2\pi e^2 n_0 r .$$

Defocusing is determined by the transverse wake. In the bubble regime, when all currents are localized in a thin layer near the bubble boundary the transverse and longitudinal wakes are related by the Panofsky-Wenzel theorem. Using Eqs. (76) and (67) one can write:

$$W_\perp(\xi,\xi_1) \approx \frac{8(\xi-\xi_1)}{r_b(\xi)^4}\theta(\xi-\xi_1) . \quad (87)$$

Assuming that initially all particles of the bunch are deflected by the same amount $r$ one obtains the wake induced force acting on particles in the bunch tail:

$$F_w \approx \frac{4e^2 N_a L_a}{r_{b1}^4} r , \quad (88)$$

where $L_a$ is the length of the accelerated bunch. Relatively simple calculations yield the following result for the ratio of deflecting to focusing strengths:

$$\eta_w = \frac{F_w}{F_r} = \frac{(r_{b2}^2 - r_{b1}^2)^2}{4 r_{b1}^4} \frac{R_b^4 + r_{b1}^2 r_{b2}^2}{R_b^4 - r_{b2}^4} . \quad (89)$$

Expressing $r_{b1}$ from Eq. (86) one obtains $\eta_w$ as a function of $\eta_P$ and $r_{b2}/R_b$. In the area of interest $r_{b2}/R_b \leq 0.5$, where acceleration is reasonably fast, $\eta_w$ can be presented by approximate formula:

$$\eta_w(\eta_P) \approx \frac{\eta_P^2}{4(1-\eta_P)^2}, \quad \frac{r_{b2}}{R_b} \leq 0.5 . \quad (90)$$

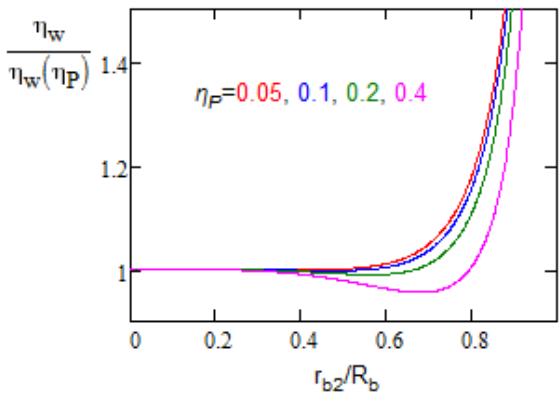

Figure 12: Ratio of $\eta_w / \eta_w(\eta_P)$ as function of $r_{b2}/R_b$.

Figure 12 presents the dependence of ratios of Eq. (89) to Eq.(90) on $r_{b2}/R_b$ for different values of $\eta_P$. One can see good coincidence $r_{b2}/R_b \leq 0.5$ and that $\eta_w$ grows fast for $r_{b2}/R_b \geq 0.5$ thus making this range of parameters uninteresting for plasma acceleration. One can see from Eq. (90) that the normalized deflection force does not depend on details of plasma acceleration and is uniquely determined by $\eta_P$ in the parameter range useful for plasma acceleration.

Strong focusing in the bubble results in a very large number of betatron oscillations in the course of beam acceleration. The total betatron phase advance can be estimated as:

$$\mu_{tot} = \sqrt{2\gamma_f} E_0 / E_a , \quad E_0 = 4\pi n_0 e / k_p ,$$

where $\gamma_f$ is the final value of particle Lorentz factor. For 1 TeV linac with $E_a/E_0=2$ it yields $\mu_{tot}/2\pi=160$. In this case, oscillations of the bunch head resonantly drive particles in the tail resulting in the increase of the effective transverse emittance.

To describe the head-tail motion we assume a rectangular longitudinal particle distribution and a linear dependence of the transverse wake on $\xi$. Then, transiting to the variables natural for betatron motion description:

$$X = \frac{x}{\sqrt{\beta_f}}\sqrt{\frac{p}{p_0}} , \qquad (91)$$

$$d\mu = \frac{dz}{\beta_f} ,$$

we obtain the equation of betatron motion:

$$\frac{d^2 X}{d\mu^2} + \frac{X}{1+\frac{\Delta p}{p}} = \frac{2\eta_w}{\left(1+\frac{\Delta p}{p}\right) L_a^2} \int_0^\xi X(\xi')(\xi-\xi')d\xi' . \quad (92)$$

Here $\beta_f$ is the beta-function determined by Eq. (20), $p_0$ is the momentum at the acceleration start, and we took into account a possible momentum deviation $\Delta p/p$ dependent on particle longitudinal position in the bunch. From a practical point of view the most interesting case is when all particles have the same initial offset, for example, excited by an error of bunch transfer between accelerating sections. For $\Delta p/p=0$ Ref. [12] presents an asymptotic solution of this problem for large $\mu$. However a solution for small $\mu$ is more important in practical estimates. Therefore, we solved the problem numerically. Figure 13 shows the dependence of growth for tail particle amplitude on $\eta_w \mu$ for different strengths of transverse wake. As one can see all presented this way solutions coincide so good that the red points on the plot are covered by blue and green ones, and blue points are covered by green ones, respectively, making them invisible for low $\eta_w \mu$. The obtained results suggest a simple approximate parameterization for amplitude of tail particle,

$$\frac{A}{A_0} = \exp\left(\frac{(\mu\eta_w)^2}{12+1.416(\mu\eta_w)^{1.57}}\right), \quad \begin{array}{l}\mu\eta_w \le 100,\\ \eta_w \le 0.1,\end{array} \quad (93)$$

and the rms amplitude of particle motion in the bunch,

$$\frac{\sigma_A}{A_0} \equiv \frac{1}{A_0}\sqrt{\frac{1}{L_a}\int_0^{L_a} A(\xi)^2 d\xi} = \exp\left(\frac{(\mu\eta_w)^2}{56+2.24(\mu\eta_w)^{1.57}}\right), \quad (94)$$
$$\eta_w \le 0.1, \quad \mu\eta_w \le 100,$$

where $A_0$ is the initial amplitude being equal for all particles. The requirement for the instability threshold depends on the requirements to the machine stability. In the case of a single small kick this requirement is more forgiving. However in a real life there are many perturbations to the machine alignment coming from ground motion, jitter in the driving beam position or positions of laser beams in the case of LPA, *etc*. Therefore, being on a safe side one needs to require that a single kick does not increase the rms motion by more than factor of 2 compared to the case when instability is absent. That yields that $\mu\eta_w \le 10$. For mentioned above 1 TeV linac we have $\mu \approx 10^3$ which yields $\eta_w \le 0.01$. Using Eq. (90) we obtain a limitation on the energy efficiency $\eta_P < 17\%$.

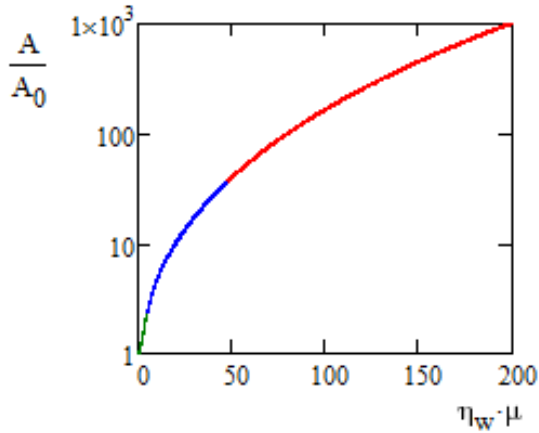

Figure 13: Dependence of the tail particle amplitude on the betatron phase advance normalized by wake strength ($\eta_w\mu$) for different strengths of the wake. Data for $\eta_w$=0.1, 0.01, 0.001 are shown by red, blue and green colors.

An effective way to suppress the BBU instability was suggested by Balakin, Novokhatsky and Smirnov [18]. It is called BNS damping. The idea is to introduce a dependence of particle momentum on the longitudinal coordinate in the bunch so that it would compensate frequency detuning due to transverse wake. In this case from Eq. (92) one obtains:

$$\frac{1}{1+\frac{\Delta p}{p}} - \frac{2\eta_w}{\left(1+\frac{\Delta p}{p}\right)L_a^2}\int_0^\xi(\xi-\xi')d\xi' = 1. \quad (95)$$

An integration and simple transformations result in:

$$\frac{\Delta p}{p} = -\eta_w \frac{\xi^2}{L_a^2}. \quad (96)$$

A requirement to have the total momentum spread of 1% yields the same value $\eta_w \le 0.01$ and, subsequently, the same limitation on the energy efficiency as for the case when BNS damping is absent. Note also that it is unclear how the quadratic dependence of momentum deviation on $\xi$ can be created on the entire length of acceleration. However one can expect some improvements of beam stability with an introduction of the energy droop along the bunch and minor improvements in the energy transfer efficiency from plasma to accelerated beam.

## 9. Other Limitations

It was already mentioned that the collapse of plasma electrons to the positron bunch center greatly affects the beam focusing coming from plasma. The focusing becomes strongly non-linear and dependent on particle longitudinal position in the bunch. All electrons which are located at radii smaller than $r_m$ (introduced in Eq. (36)) are pulled into the positron bunch. Even for a modest number of positrons in a bunch, $r_m$ is orders of magnitude larger than the transverse beam size. Note that the beam emittance is fixed by the luminosity and the only way to increase the beam size is an increase of beta-function but it is limited by the necessity to prevent the emittance growth due to multiple scattering and to suppress the BBU instability. For a bunch population of $4\cdot 10^9$ and a bunch length of 10 μm one obtains $r_m$=10.6 μm. The frequency of plasma electrons oscillations increases inversely proportional to their impact parameters. It results in good mixing of plasma electrons near axis making variations of their density along the axis smooth. In the absence of electron interaction their density near axis would diverge logarithmically with radius. A repulsion of electrons prevents divergence but still cannot suppress very large electron density variation and a plasma focusing nonlinearity related to it. Such nonlinear focusing would not create tremendous problems with emittance growth in the case of a single accelerating section and an absence of the final focus. But it is not the case for a plasma-based collider. Another way to characterize the problem is the phase advance of small amplitude oscillations of plasma electrons in the field of positron bunch:

$$\mu_e = \frac{\sqrt{2r_e N_a L_a}}{\sigma_\perp} = \sqrt{2}\frac{r_m}{\sigma_\perp}. \quad (97)$$

Here $\sigma_\perp$ is the rms transverse size of positron bunch and a round beam is assumed. For considered above parameters ($N_a$=4·10$^9$, $L_a$=10 μm) and $\sigma_\perp$=0.15 μm (see Figure 1) one obtains $\mu_e$=100. To have reasonably linear focusing this phase advance need to be reduced by at least two orders of magnitude. That requires a drastic decrease for number of positrons per bunch.

In the case of electron bunch acceleration we have a problem of collapsing plasma ions in the field of electron bunch. In this case in the expression for $r_m$ one needs to use the ion mass instead of the electron one. For above considered case of electron beam acceleration with the bunch population of $8.86 \cdot 10^9$ and the bunch length of 4.2 μm one obtains $r_m$=0.2 μm for proton plasma. This size is still larger than the electron beam radius varying in the range of 0.05 – 0.15 μm (see Figure 1). That means that a problem of ion collapse in the field of electron bunch is also quite severe and will be an important limitation on the collider parameters. Similar to the case of positrons we introduce the phase advance of small amplitude oscillations of plasma ions in the field of electron bunch:

$$\mu_{ion} = \frac{\sqrt{2 r_p N_a L_a}}{\sqrt{M_i} \sigma_\perp} ,$$

where $M_i$ is the ratio of the ion mass to the proton mass. For discussed above parameters of the accelerated electron bunch and proton plasma the phase advance grows with beam energy and achieves ~360° at 500 GeV. To avoid problems this value needs to be reduced by at least an order of magnitude.

Note that although the use of heavy ions looks as a possible means to mitigate the problem of ion collapse it does not look as a real possibility due to the impact ionization of the ions. For the bunch parameters mentioned in the previous paragraph one obtains the electric field at the bunch boundary exceeding $10^3$ GV/cm. It is more than two orders of magnitude larger than the electric field in a hydrogen atom of ~6 GV/cm. Using ions stripped to the level sufficient to avoid impact ionization looks to be un-realistic. Heavy ions also increase the effects of bremsstrahlung, which are not completely negligible even for proton plasma.

## Discussion

There are a number of problems which need to be resolved before a credible concept of plasma-based e$^+$-e$^-$ or γ-γ collider can be put forward. In this paper we intentionally discuss only beam physics limitations leaving aside multiple outstanding challenges in technology and engineering. As far as we can presently judge there is still no viable path, which could lead to a high luminosity collider concept within the present paradigm.

The most outstanding problem is the acceleration of positrons with bunch brightness, required for a linear collider. One needs to have plasma electrons at the system axis to have sufficiently strong focusing required to suppress the transverse BBU (hosing) instability. The plasma transverse impedance (transverse wake) is very large and the suppression of this instability cannot be done by means other than plasma focusing. Any other existing focusing is orders of magnitude weaker than the plasma focusing and is incapable of suppressing the instability. But the problem is that the presence of plasma electrons at the positron path results in their pinching by the bunch head. It creates a very non-linear focusing field in the tail. This field is driven by the high density of plasma electrons at the axis which is orders of magnitude higher than the initial electron density in the plasma. This high density of electrons also greatly increases multiple scattering on the plasma electrons, resulting in unacceptably large emittance growth. The plasma channel was suggested to mitigate pinching and multiple scattering. In this case there are no charged and neutral particles in close vicinity of system axis. Consequently, such an arrangement solves problems of multiple scattering and pinching. However it brings another problem - the absence of mechanism capable to suppress the BBU instability due to the large transverse impedance of a channel. To address it, one needs external focusing orders of magnitude higher than can be achieved with conventional means.

The situation is much better in the case of electrons. Potentially, one can consider a γ-γ or e$^-$-e$^-$ collider but its luminosity will still be significantly lower than the luminosity of ILC or CLIC. There are several main phenomena which limit the luminosity. All of them, one way or another, are related to the efficiency of energy transfer to the beam. Studies of luminosity limitations carried out for the ILC and CLIC show that in the case of a global machine optimization, the collider luminosity depends only on the beam power; and that an achievement of luminosity, comparable to the LHC luminosity, requires the beam power above tens of megawatt. The operation at this power levels requires very high efficiency of energy transfer to the accelerated beam. In the ILC and CLIC it is achieved by acceleration of a very large number of bunches in a single linac pulse. Such an operation is supported by high Q-values of accelerating cavities. In this case, each bunch after passing a cavity receives a small percentage of energy stored in this cavity. The lost energy is replenished by the RF power source before the arrival of the next bunch. As result, in the case of ILC, after passing many bunches, about 20% of RF energy is transferred to the beam. Consequently, the overall energy efficiency is about 10%.

Achieving such an efficiency in the plasma-based accelerator represents a great challenge. The problem originates from a low Q-value of plasma oscillations (especially in the bubble regime), resulting in that only one bunch can be accelerated in a single pulse. Contrary to conventional accelerating structures, a plasma based accelerating structure is excited at a single frequency which is close to the plasma frequency. Consequently, an absence of higher order modes suggests that the efficiency of plasma acceleration for a single bunch can be significantly higher than for a conventional structure. Although this statement is supported by a number of simulations carried out in recent years, we are still far away from demonstrating that it could work for acceleration of a collider quality beam. The above mentioned BBU instability, driven by transverse impedance of the plasma bubble, is one of the major problems. It limits the number of particles in the bunch and, consequently, limits the efficiency of acceleration. In optimistic scenarios, the instability does not lead to unacceptably large emittance growth in the course of acceleration, however it still greatly amplifies the emittance growth due to errors of the relative alignment of different accelerating sections. Note that presently the required alignment accuracy does not look achievable even in the absence of the BBU instability. The BNS damping, which potentially could help, requires too large of an energy spread, which is inacceptable from the final focus point of view.

For present proposals, pinching of plasma ions by a bright electron beam limits the luminosity of $e^--e^-$ or $\gamma-\gamma$ collider to well below $10^{34}$ cm$^{-2}$s$^{-1}$ if a plasma with light ions is used. Using heavy ions, which cannot be completely stripped due to required energy efficiency, is prohibited by their impact ionization by an electron bunch fields. It also greatly amplifies multiple scattering and bremsstrahlung.

In conclusion, we would like to stress that there are many potential application for plasma-based accelerators. However, presently it is unclear how the above mentioned limitations could be overcame for high luminosity linear collider.